\RequirePackage{fix-cm}
\documentclass[twocolumn,epjc3]{svjour3}
\smartqed  
\RequirePackage{graphicx}
\RequirePackage[caption=false]{subfig}
\RequirePackage{xcolor}
\usepackage{bm}
\usepackage{xspace}
\usepackage{multirow}
\usepackage{xcolor}
\usepackage[british]{babel}
\usepackage{hyphenat}
\usepackage{amsmath,amssymb}

\newcommand{\beq}{\begin{equation}}
\newcommand{\eeq}{\end{equation}}

\def\Xmax{$X_{\rm max}$\xspace}
\def\XmaxZero{$X_{\rm max}^{0}$\xspace}
\def\XmaxOne{$X_{\rm max}^{1}$\xspace}
\def\Sem{$S_{\rm em}$\xspace}

\journalname{Eur. Phys. J. C}
\begin{document}

\title{New methods to reconstruct $X_{\rm max}$ and the energy of gamma-ray air showers with high accuracy in large wide-field observatories} 


\author{R. Concei\c{c}\~ao\thanksref{addr1,addr2}
        \and
        L. Peres\thanksref{addr1, addr2}
        M. Pimenta\thanksref{addr1,addr2}
        \and
        B. Tom\'e\thanksref{addr1,addr2}
}

\thankstext{e1}{e-mail: ruben@lip.pt}


\institute{
LIP, Av. Prof. Gama Pinto, 2, P-1649-003 Lisbon, Portugal \label{addr1}
\and
Instituto Superior T\'{e}cnico (IST), Universidade de Lisboa, Av. Rovisco Pais 1, 1049-001, Lisbon, Portugal \label{addr2}
}

\date{Received: date / Accepted: date}

\maketitle

\begin{abstract}

Novel methods to reconstruct the slant depth of the maximum of the longitudinal profile (\Xmax) of high-energy showers initiated by gamma-rays as well as their energy ($E_0$) are presented. The methods were developed for gamma rays with energies ranging from a few hundred GeV to $\sim 10$ TeV.
An estimator of \Xmax is obtained, event-by-event, from its correlation with the distribution of the arrival time of the particles at the ground, or the signal at the ground for lower energies.
An estimator of $E_0$ is obtained, event-by-event, using a parametrization that has as inputs the total measured energy at the ground, the amount of energy contained in a region near to the shower core and the estimated \Xmax.

Resolutions about $40 \, (20)\,{\rm g/cm^2}$ and about $30 \, (20)\%$ for, respectively, \Xmax and $E_0$ at $1 \, (10) \ \rm{TeV}$ energies are obtained, considering vertical showers.
The obtained results are auspicious and can lead to the opening of new physics avenues for large wide field-of-view gamma-ray observatories. The dependence of the resolutions with experimental conditions is discussed.

\end{abstract}

\section{Introduction}
\label{sec:intro}

High energy cosmic and gamma rays entering the Earth atmosphere originate Extensive Air Showers (EAS) \sloppy \ which may be characterised by the distributions of the number of shower particles $N$ as a function of the traversed atmospheric slant depth $X$ (longitudinal profiles) and/or by the distributions of the particles arriving at the ground level as a function of the distance to the shower core (Lateral Density Function - LDF).

The longitudinal development of gamma-ray initiated showers was historically described by Rossi and Greisen diffusion equations \cite{Rossi-Greisen} being the well known Greisen \cite{Greisen} and  Gaisser-Hillas \cite{Gaisser-Hillas} functions approximate solutions. It can be demonstrated that these functions lead to a quasi-universal shape \cite{Longi_profiles}. This universality can be shown by representing the shower longitudinal profile in the plane ($N^{\prime} = N/N_{\rm max}$ ,  $X^{\prime} = X - X_{\rm max}$), where \Xmax is the slant depth of the maximum of the profile and $N_{\rm max}$ is the number of the shower particles at that depth. In this reference frame, the profile may be seen as a slightly asymmetric Gaussian with variable width and is essentially insensitive to variations induced by the depth of the first interaction \cite{RL}. However, at TeV energies or below, the fraction of events where the longitudinal profile do not follow the quasi Gaussian shape may not be negligible. A few of the profiles will have a slower decrease after \Xmax or even having a double peak structure. These anomalous shower profile structures are associated with interactions where particles travel several radiation/interaction lengths before interacting, or when one of the sub-products of the interaction takes nearly all of the available energy.

Imaging Air Cherenkov Telescopes (IACTs) collect Cherenkov light produced by the EAS and are able, due to their size, intensity, and orientation of the projected image in the camera focal plane, to reconstruct the energy and the direction of the primary gamma-ray.
Typical energy resolution of $15\%-20\%$ are often quoted for TeV gammas and zenith angles of about $20^{\circ}$, for instance, the resolution of MAGIC-II was measured to be 15\% \cite{MAGIC}.

Ground-based gamma-ray observatories sample the particles (mainly electrons and photons) arriving at the ground level and from their time and position distributions can determine, with reasonable accuracy, the shower core position and the direction of the primary gamma-ray. The determination of the shower energy has, however, a large uncertainty. Indeed, energy resolutions of the order of several tens of percent are often quoted. For instance, in~\cite{Vikas} using a Likelihood fit with Monte Carlo template distributions, the energy reconstruction resolution at $10\,$TeV is $\sim 50\%$ and the HAWC collaboration has recently reported an improved energy resolution of $40\%$ at the same energy using a neural network analysis~\cite{HAWC_energy}.

One of the main limiting factors in the reconstruction of the primary energy for ground arrays is the uncertainty on the position of the first interaction in the atmosphere. Contrary to IACT arrays, there is no direct measurement for ground arrays of the contents of the EAS in the region of the shower maximum, and therefore the shower development stage is unknown. In fact, for showers induced by gamma-rays, with the same energy and zenith angle, the number of particles at the ground is expected to increase with \Xmax. The previous statement is not absolute as the width of the longitudinal profile is also an important factor to determine not only the total energy at the ground but also the fraction of this energy present in the region of the shower core. A larger shower profile width will, for the same \Xmax,  have larger energy at the ground and a more substantial fraction of energy in the region near the core. Indeed,  it is possible to establish, at fixed primary energy, a correlation between the fraction of the energy in the region near the core and the total energy carried by electromagnetic particles (photons, electrons and positrons) that reach the ground (\Sem).

The energy  measured by a detector array with  electromagnetic calorimetric capabilities, like Water Cherenkov Detectors, is by definition highly correlated with \Sem . Such correlation is  explored  to built an estimator of \Sem (section \ref{sec:ShowerStage}).


\Xmax is not easily estimated at ground gamma-ray observatories.
However, at very high energies in cosmic rays experiments, correlations between \Xmax and the distribution of the arrival times of the particles at the ground in each event, have been established \cite{SD_XMAX}.
Such correlations are exploited to build an estimator of \Xmax  (section \ref{sec:Xmax}).

An innovative method for the reconstruction of the primary energy of each event, having as inputs the total measured energy at the ground,  the fraction of this energy measured in the region near the shower core and the estimated \Xmax, is then presented (section \ref{sec:Erec}).

Finally, the applicability of such method in realhistogrammed
large wide-field gamma-ray observatories is discussed (section \ref{sec:conclusions}).

All the present results were obtained using CORSIKA (version 7.5600) \cite{CORSIKA} to simulate vertical gamma-ray showers assuming an observatory at an altitude of 5200 m a.s.l. The primaries, with energies between $250\,$GeV and $15\,$TeV, were injected following an energy spectrum of $E^{-1}$, which guarantees high enough statistics over the whole simulated energy range. It was used as hadronic interaction model for low and high energy, FLUKA and QGSJET-II.04, respectively, although the choice of these models has little impact on the simulation of electromagnetic showers. The total energy of electromagnetic shower particles was recorded at the observation level and histogrammed in radial bins of 4 meters. This would mimic a calorimeter detector compact array, where the station unit covers an area of $\sim 12\,{\rm m^2}$.
A full study including different zenith angles,  detailed simulations of realistic detectors, a wider energy range and also its application to hadronic induced showers, is out of the scope of this work. The main focus of this article is the explanation of the newly proposed method to achieve an accurate energy reconstruction of gamma-ray air showers between several hundreds of GeV and a few TeV. This is the most challenging energy region for this kind of experiments. In addition to the uncertainty on the shower stage, the number of secondary shower particles that reach the ground is reduced when comparing with energies of the order of tens of TeV, undermining the capability to reconstruct the shower quantities.
A first glance on its potential is given.

\section{Energy distribution at the ground }
\label{sec:ShowerStage}

The density of the shower particles arriving at the ground,  as a function of the distance to the shower core, is steeper in the region near the core and flatter at larger distances.
This distribution is usually parametrized using the NKG  (Nishimura-Kamata-Greisen) formula \cite{NKG}.
The particle density at a given distance from the air shower axis is often used to obtain an estimator of the primary energy, trying to find a region less sensitive to the fluctuations in the shower development, to the primary nature and to array sampling effects. For instance, in the Tibet Air Shower Array, this distance was found to be 50 m \cite{tibet}.

\begin{figure}[!t]
  \centering
  \includegraphics[width=0.5\textwidth]{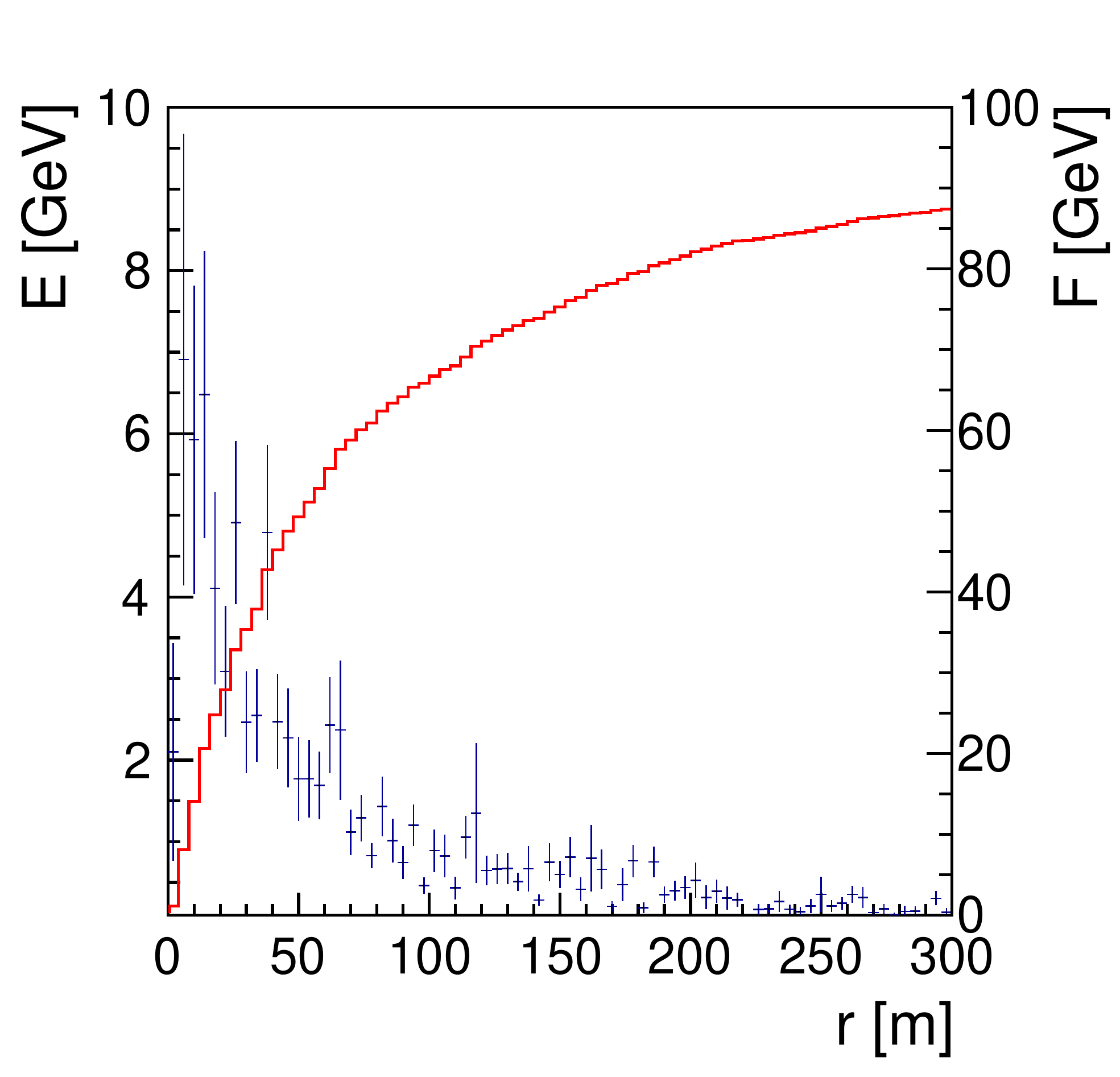}
  \caption{\label{fig:E_ground} The energy distribution at the ground for one event with $S_{\rm em} = 96.5\,$GeV, $E_0= 1165.9\,$GeV and \Xmax $= 334 \ \rm{g\,cm^{-2}}$. Also shown the respective  cumulative function $F(r)$.}.
\end{figure}

 The energy distribution at the ground as a function of the distance $r$ to the core position and its cumulative function, $F(r)$, are shown in figure \ref{fig:E_ground} for an event with \Sem $ = 96.5\,$GeV, $E_0= 1165.9\,$GeV  and \mbox{\Xmax $= 334 \ \rm{g\,cm^{-2}}$}.

The strategy followed in this article is different from one usually employed by shower array experiments~\cite{HAWC_energy,LHAASO_energy}. Instead of using the shower particle density at the ground at some optimized distance from the shower core, the aim is to characterize the shower development through two variables that will be then used  to predict, event by event, the calibration factor between the gamma-ray energy ($E_0$) and the electromagnetic energy arriving at the ground (\Sem).

\Xmax, whose estimator will be discussed in the next section, is naturally one of these variables. The other, $f_{20}$, is defined as the ratio between the energy at the ground collected at a distance less than $20\,$m from the shower core and the total energy at the ground. This parameter will be the main responsible for the improvement of the energy resolution reached in this article. The rationale of this second variable was already introduced in the previous section. For a given $E_0$ and \Xmax, the development of the shower between the \Xmax region and the ground level will strongly determine $f_{20}$.

 An estimator of \Sem, designated as $A_0$, may be obtained using  the correlation between \Sem  and \mbox{$F_{r_0} \equiv F(r_0)$}, being $r_0$ a reference distance. In any case, $r_0$ should be greater than $20$ m to ensure a good correlation, and lower than some tens of meters to ensure a high number of events where the event footprint, with  $ r \, < r_0$, is fully contained within the compact array region of the observatory.
 For the purpose of this article, in the following $r_0 = 50\,$m is used.

 The correlation between \Sem and $F_{50}$ is shown in figure \ref{fig:F50Sem} and $A_0$ is parametrized as:

\begin{figure}[!t]
  \centering
  \includegraphics[width=0.5\textwidth]{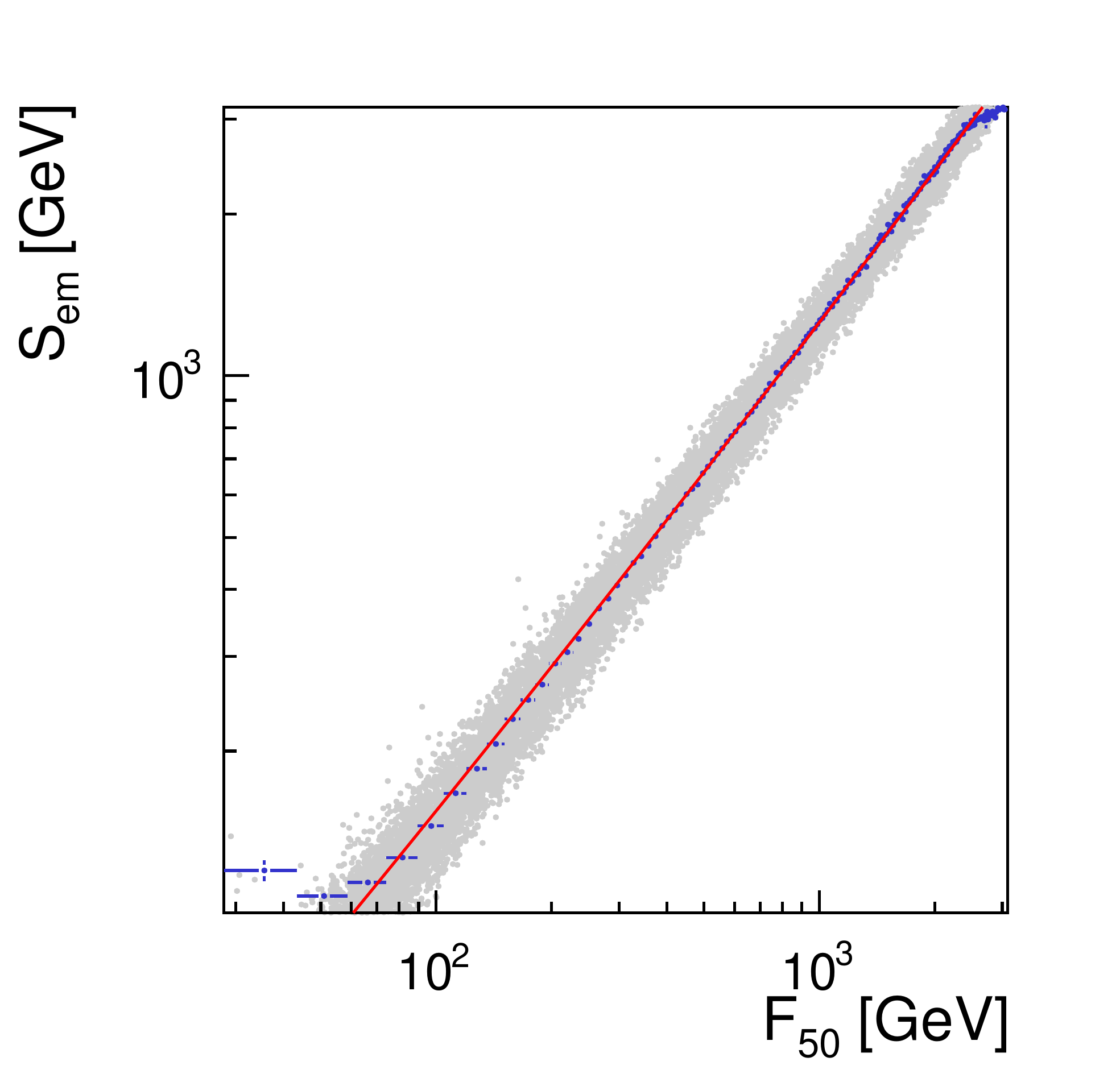}
  \caption{\label{fig:F50Sem} Correlation of the electromagnetic energy deposited at the ground, \Sem,  with $F_{50}$, the energy deposited at the ground within a radius  of 50 m from the core position (see text for the simulation details).}
\end{figure}

\begin{center}
\begin{equation} \label{eq:A0}
A_0 \, = \,F_{50}\, +\, G \, F_{50}^{\delta}  \,
\end{equation}
 \end{center}

\noindent where $G$ and $\delta$ are free positive parameters. This parametrization ensures, by construction,  that $A_{0}$ is always greater then $F_{50}$.
With $A_{0}$ and $F_{50}$ in GeV, the best values found to be for $G = 1.63\,{\rm GeV}^{0.28}$ and $\delta = 0.72$.
The result is shown as a red curve in
figure \ref{fig:F50Sem}.

\begin{figure}[!t]
  \centering
  \includegraphics[width=0.5\textwidth]{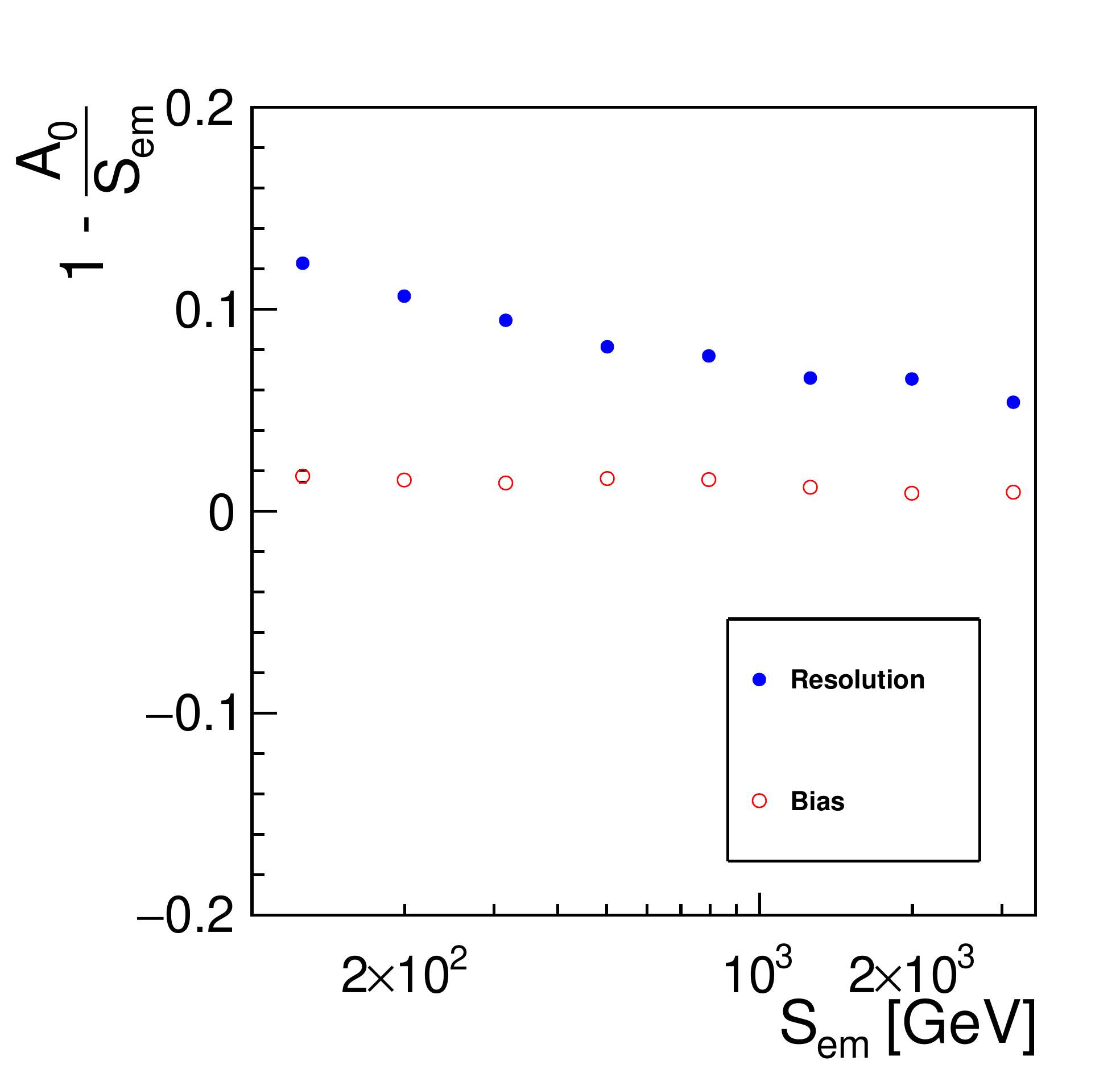}
  \caption{\label{fig:A-resol} Bias and the relative resolution of the estimator $A_0$ of the energy deposited at the ground,  defined in equation \ref{eq:A0}, as a function of the energy deposited at the ground, \Sem(see text for the simulation details).}
\end{figure}

The obtained resolutions and bias\footnote{In this work, the bias and resolutions of the estimator, $\hat{x}$, of variable, $x$, are taken fitting a gaussian function to the residuals, ($1-(\hat{x}/x)$). The bias and the resolution corresponds to the mean and the sigma parameter of the fitted gaussian, respectively.} of $A_0$  are summarized in figure \ref{fig:A-resol}, as a function of \Sem.
As a reference a primary energy of $1\,$TeV and  $10\,$TeV corresponds to a mean value of \Sem  of $115\,$GeV and $3\,$TeV, respectively (see figure \ref {fig:E0-SemvsSem}). Thus, resolutions of about $12 \%$ and $5 \%$
are found at primaries energies of $1\,$TeV and  $10\,$TeV, respectively, while the  bias is consistently in the order of a few percent.

In principle, a better estimator of \Sem can be obtained using a continuously measured distribution of $F(r)$ and not just a single measurement at $r_0$. With this purpose,
$F(r)$, which is a smooth and continuous function, was parametrized as:

\begin{center}
\begin{equation} \label{eq:par}
F(r)= A_1\left[1 - \exp\left(-\frac {k_1 r^{\alpha_1}} {1+k_2 r^{\alpha_2}}\right) \right].
\end{equation}
 \end{center}

The parameter $A_1$ is the \Sem estimator  while the terms ${k_i r^{\alpha_i}}$ describe the steepness of the function, being
$k_{1} >0,\, \alpha_{1} > 0 $ and $0 \leq k_{2} \leq 1, \,  0 \leq \alpha_{2} < \alpha_{1} $.

In the limits $ r \rightarrow 0 $ and $ r \rightarrow \infty $ this parametrization becomes,

\begin{center}
\begin{equation}
F(r)/A_1 = \left[1 - \exp\left(- {k_1 r^{\alpha_1}} \right) \right]\,,
\end{equation}
 \end{center}

 and,

 \begin{center}
\begin{equation}
F(r)/A_1= \left[1 - \exp\left(- {\frac{k_1} {k_2} \, r^{(\alpha_1- \alpha_2)}} \right) \right],
\end{equation}
 \end{center}

\noindent which have the form of  Weibull cumulative distribution functions \cite{Weibull}.

\begin{figure}[!t]
  \centering
  \includegraphics[width=0.5\textwidth]{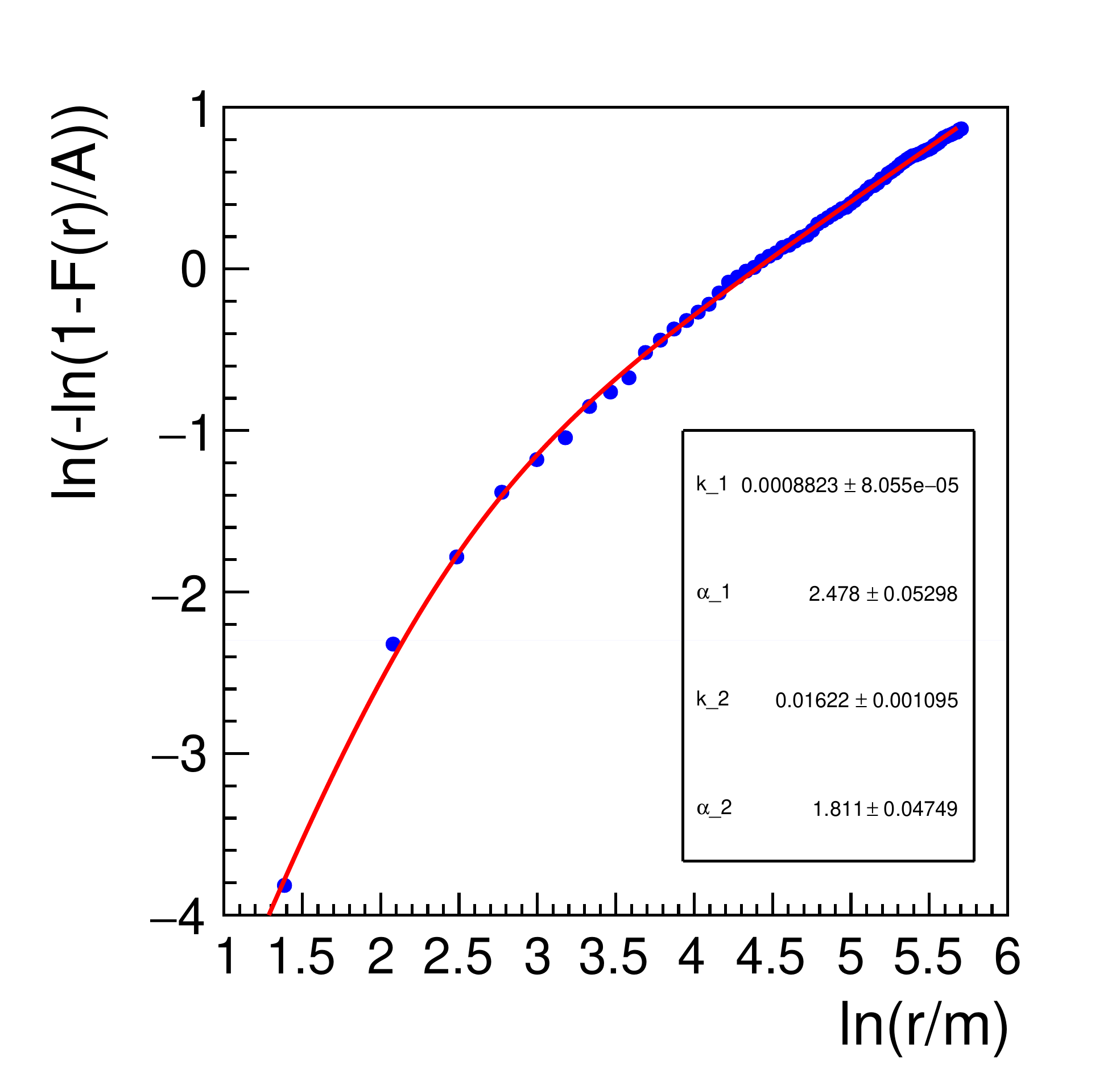}
  \caption{\label{fig:lnln(1-fr)} The blue points represent the binned cumulative function of the electromagnetic energy distribution at the ground  of the event shown  in figure~\ref{fig:E_ground}, assuming $A_1 = S_{\rm em}$. The red line is the best fit to these data points using the parametrization defined in equation \ref{eq:par}.
 }.
\end{figure}

Indeed, as an example, it is shown in figure \ref{fig:lnln(1-fr)} that, in the plane  ($\ln(-\ln(1-(F(r)/A_1)))$,  $\ln(r)$) and assuming $A_1 = S_{\rm em}$,
the cumulative distribution function of the event shown  in figure~\ref{fig:E_ground}  (blue points)  is well described by the above parametrization.



In this plane a pure Weibull function is just a straight line while to describe this event two straight lines are
needed due the transition between the two regimes in the region $20\,\mathrm{m} - 30\,\mathrm{m}$ (\mbox{$\ln(r/m) =$ 3 - 3.5}). This transition region matches the known behaviour of the energy distribution at the ground with a higher concentration of energy in the core region. Similar fits to a large number of  events and whose cumulative functions correspond to quite different steepness in the core region are equally very good.

  In a real event  \Sem is not known and the parameter $A_1$, as well as the parameters $k_{1}$, $\alpha_{1}$, $k_{2}$ and  $\alpha_{2}$, have to be fitted.
However, it was found that the convergence of this fit is not trivial, as  $A_1$ is highly correlated with combinations of the other parameters, and thus elaborated fit strategies will have to be defined, which is beyond the scope of the present article.
In these terms $A_0$ will be the \Sem estimator used hereafter.

The variable $f_{20}$ is then defined as $F_{20}/A_0$.
The choice of 20 m for the definition of this variable is a compromise which should be optimized for each specific experiment. Nevertheless, its value should be typically between 15 m and 30 m. Lower values will conflict with the possible experimental resolutions on the shower core, higher values will enter in the region where the cumulative function has a slower increase and also where, for events with the core nearer to the border of the compact region of the array, there will be no direct measurement of the cumulative function.

\section{\Xmax reconstruction and resolution}
\label{sec:Xmax}

\begin{figure}[!t]
  \centering
  \includegraphics[width=0.5\textwidth]{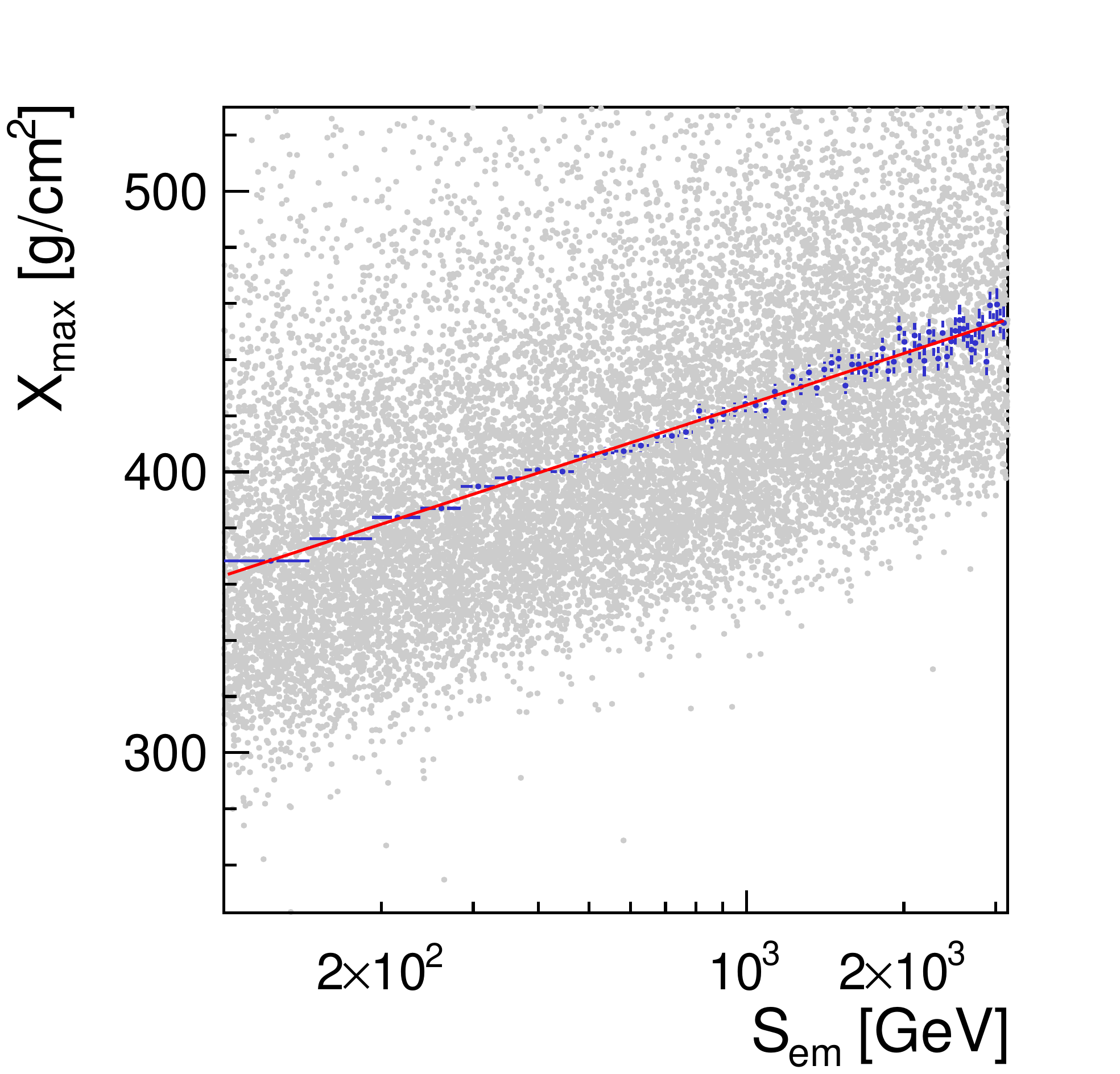}
  \caption{\label{fig:XmaxvsSem}  Correlation of the slant depth of the maximum of the air shower longitudinal profile, \Xmax,  with the energy deposited at the ground, \Sem (see text for the simulation details).}
\end{figure}

A first order estimation of \Xmax may be obtained observing that the mean value of \Xmax increases with the increase of the electromagnetic energy arriving at the ground (\Sem), reflecting the increase of the shower size with the primary energy. This correlation is demonstrated in figure  \ref{fig:XmaxvsSem} where \Xmax is represented as a function of \Sem.
It is then possible to parametrize \Xmax as a function of \Sem as:

\begin{center}
\begin{equation}
X_{\rm max}^{0} \, = B_0 +\, \gamma_0 \, \log(S_{\rm em}/\mathrm{GeV}).
\end{equation}
\end{center}

\noindent with $B_0$ and $\gamma_0$ parameters tuned to describe the mean behaviour.
The best achieved parameterization is shown in
figure \ref{fig:XmaxvsSem} as a red filled curve (corresponding to
$B_0 = 237.1 \, \mathrm{g\,cm^{-2}}$
and
$\gamma_0 = 62.3 \, \mathrm{g\,cm^{-2}}$
).

\parskip 1.5ex

A more precise estimate of \Xmax may be obtained exploring the fact that the shower front at the ground is a curved surface~\cite{showerCurvature}. Ideally, if the shower particles were produced in a single point located along the shower axis, for instance at the \Xmax, this surface will be spherical, assuming that all the particles travel approximately with the speed of light.
The arrival time in each surface station would then change accordingly as a function of the distance to the shower core and with the primary particle direction. Using a simple geometrical fit, the \Xmax position would be reconstructed straightforwardly,  with an accuracy that would mainly depend on the time resolution of the stations.

\parskip 0 ex

\begin{figure}[!t]
  \centering
  \includegraphics[width=0.5\textwidth]{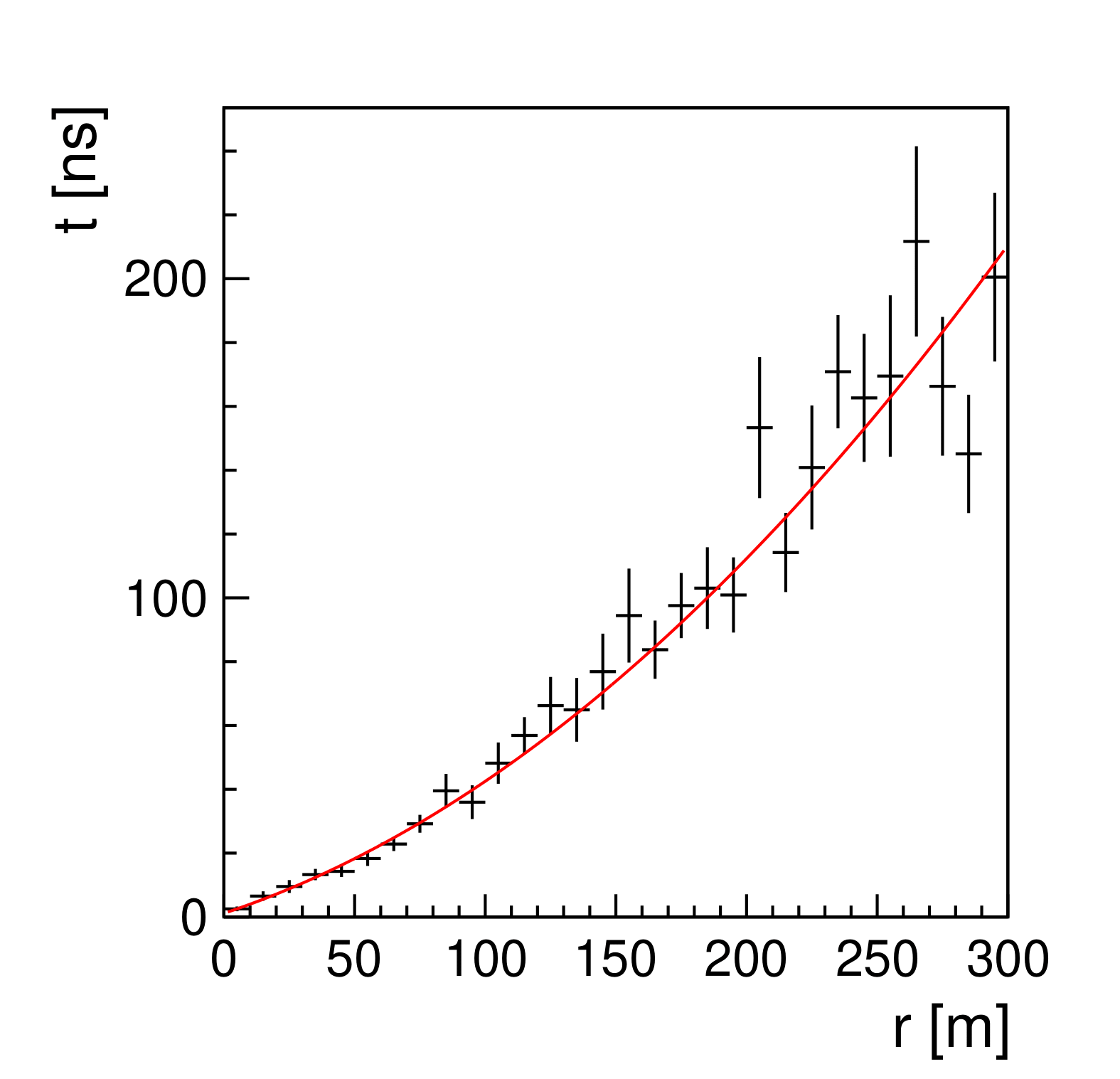}
  \caption{\label{fig:fitcurvature} Arrival time as a function of the distance to the core for an event with $X_{\rm max} = 339 \,\rm{g\,cm^{-2}}$ and $E_0=1.3\,$TeV. The line correspond to a quadratic fit as explained in the text with its $\chi^2/{\rm ndf} = 0.93$ (see text for the simulation details). }.
\end{figure}

In reality, the geometry is more complex, but nevertheless, it is possible to establish a clear correlation between \Xmax  and the arrival time distribution of particles at the ground.
As an example in figure \ref{fig:fitcurvature} this correlation is shown for an event with $X_{\rm max} = 339 \, \rm{g\,cm^{-2}}$ and $E_0= 1.3\,$TeV.
It was found that most of the events can be described by a quadratic polynomial of the form,

\begin{center}
\begin{equation} \label{eq:delta}
t = a + b \, r + c \, r^2.
\end{equation}
\end{center}

In fact, the application of the above equation to the time profiles as a function of the distance to the shower core leads to a well behaved $\chi^2/ndf$ distribution with the distribution maximum peaking at $\sim 1.2$.

\begin{figure}[!t]
  \centering
  \includegraphics[width=0.5\textwidth]{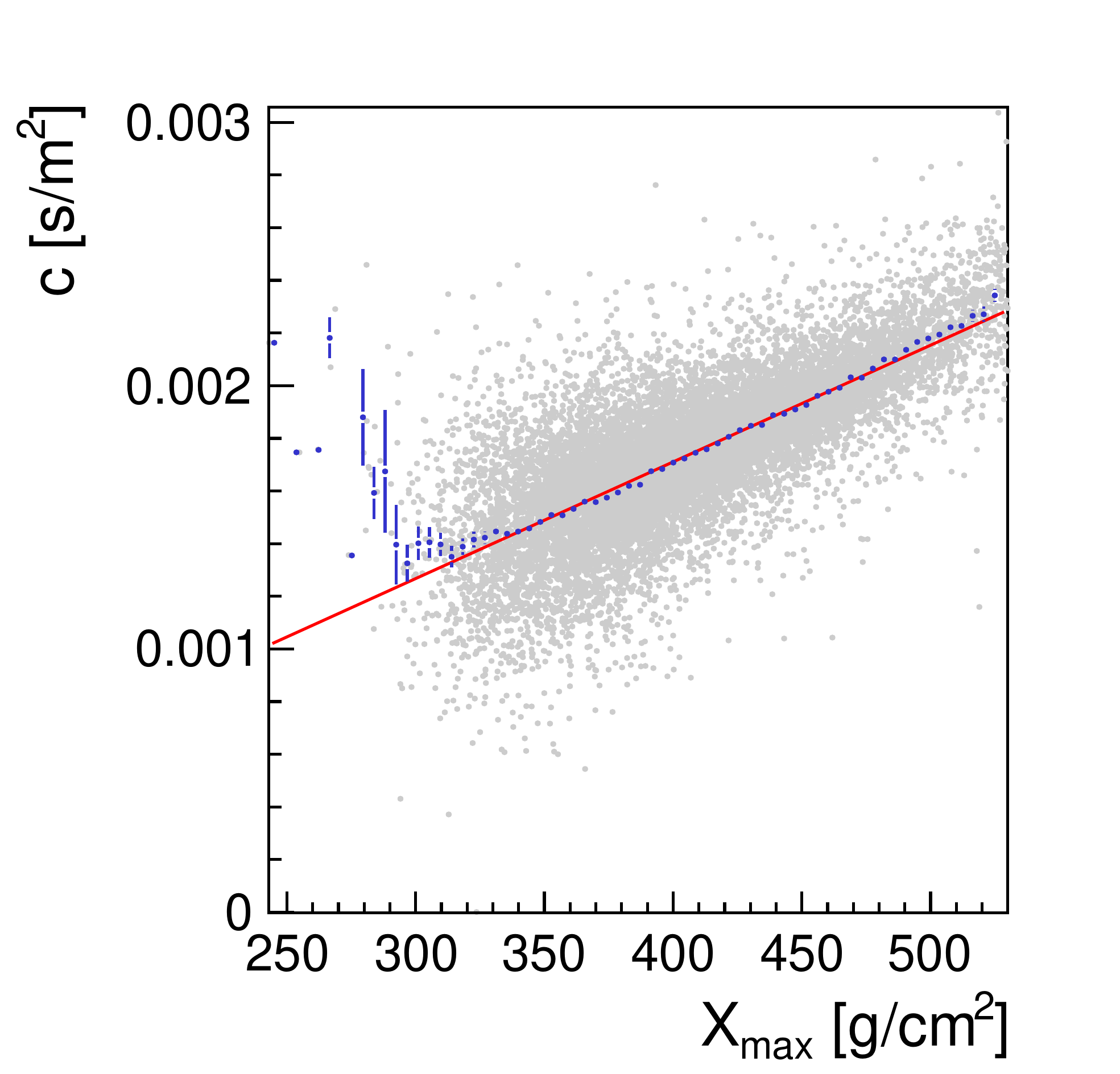}
  \caption{\label{fig:XMAX_c} Correlation of  the $c$ curvature parameter with the slant depth of the maximum of the longitudinal profile, \Xmax. See equation \ref{eq:delta} for a definition of the $c$  parameter (see text for the simulation details).}
\end{figure}

 The parameter of the quadratic term of the polynomial, $c$, is strongly correlated with \Xmax (figure \ref{fig:XMAX_c}). The parameter $b$ is nearly independent of \Xmax, and $a$ is associated with the event initial time, $T_0$, usually set to zero when the shower front reaches the shower core position.
 The dependence of $c$ with \Xmax can be understood if one assumes that most of the particles produced in a shower come from \Xmax and the shower particles propagate as spherical front. This is of course an approximation but figure~\ref{fig:XMAX_c} supports it and it helps to build some intuition.

 Hence, it is possible to parametrize \Xmax as a function of $c$ using:

\begin{center}
\begin{equation}
X_{\rm max}^{1} \, = B_1 +\, \gamma_1 \, c.
\label{eq:Xmax_c}
\end{equation}
\end{center}

$B_1$ and $\gamma_1$  are parameters tuned to describe the profile shown in figure~\ref{fig:XMAX_c}.
The best achieved parametrization is shown by the red curve, with
$B_1 = 11.2\, {\rm g\,cm^{-2}}$ and
$\gamma_1 = 2.28\times10^{9}\, {\rm g\,s^{-1}}$.

\begin{figure}[!t]
  \centering
  \includegraphics[width=0.5\textwidth]{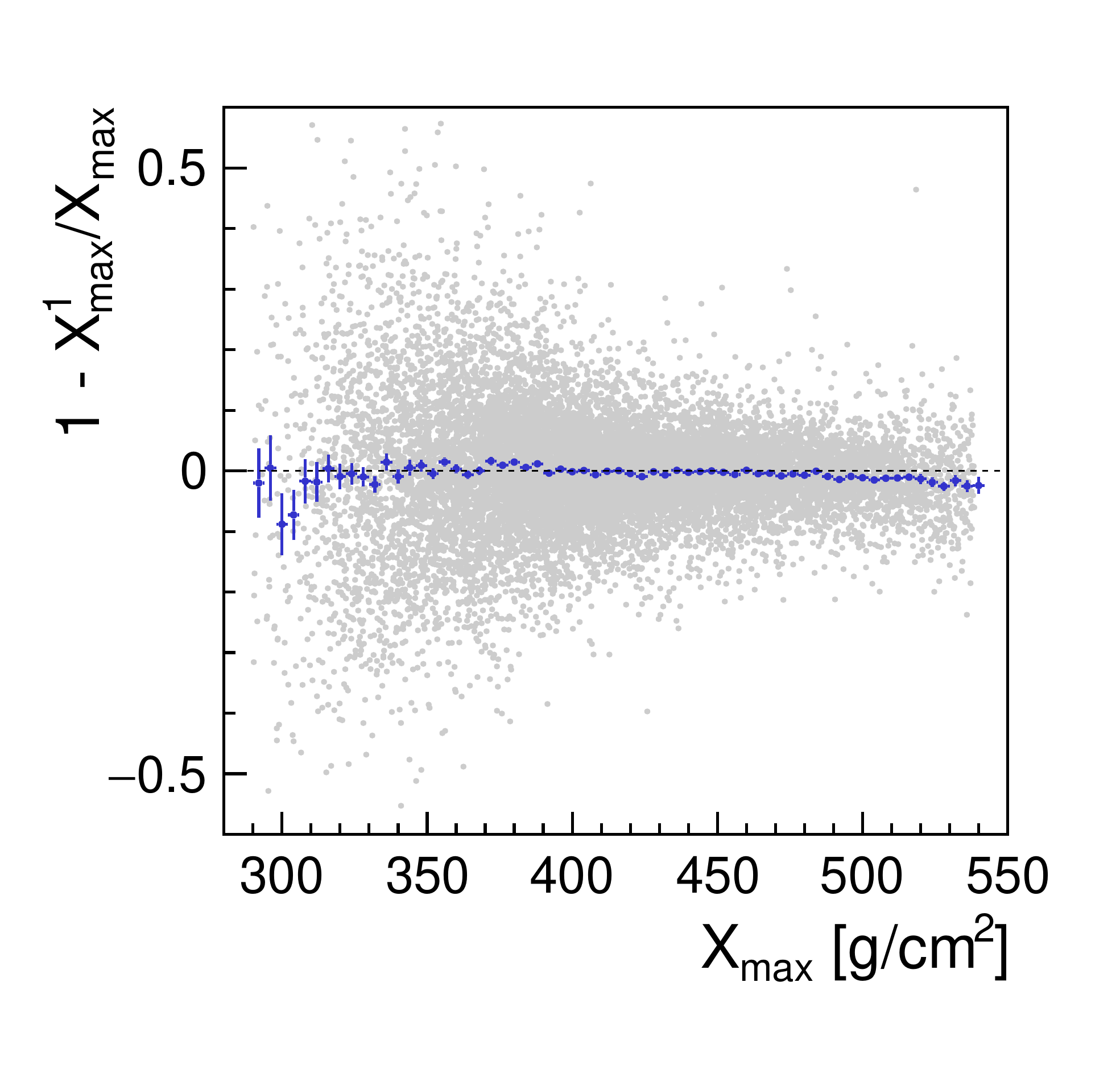}
  \caption{\label{fig:XMAX-rec} Normalized deviation of the estimator \XmaxOne, defined in equation \ref{eq:Xmax_c}, from the real \Xmax,  as a function of \Xmax (see text for the simulation details).}
\end{figure}

\XmaxOne does not show any relevant bias even for low \Xmax, as shown in figure \ref{fig:XMAX-rec}.
Nevertheless, in a few cases, particularly at lower energies where the number of particles arriving at the ground is small, the fit may converge to $c$ values leading to non-physical values of \Xmax. In practice, whenever the estimation of \Xmax from the fit
indicates values lower than $300 \ \rm{g\,cm^{-2}}$, the first order estimation \XmaxZero  is used.

\parskip 0 ex

\begin{figure}[!t]
  \centering
  \includegraphics[width=0.5\textwidth]{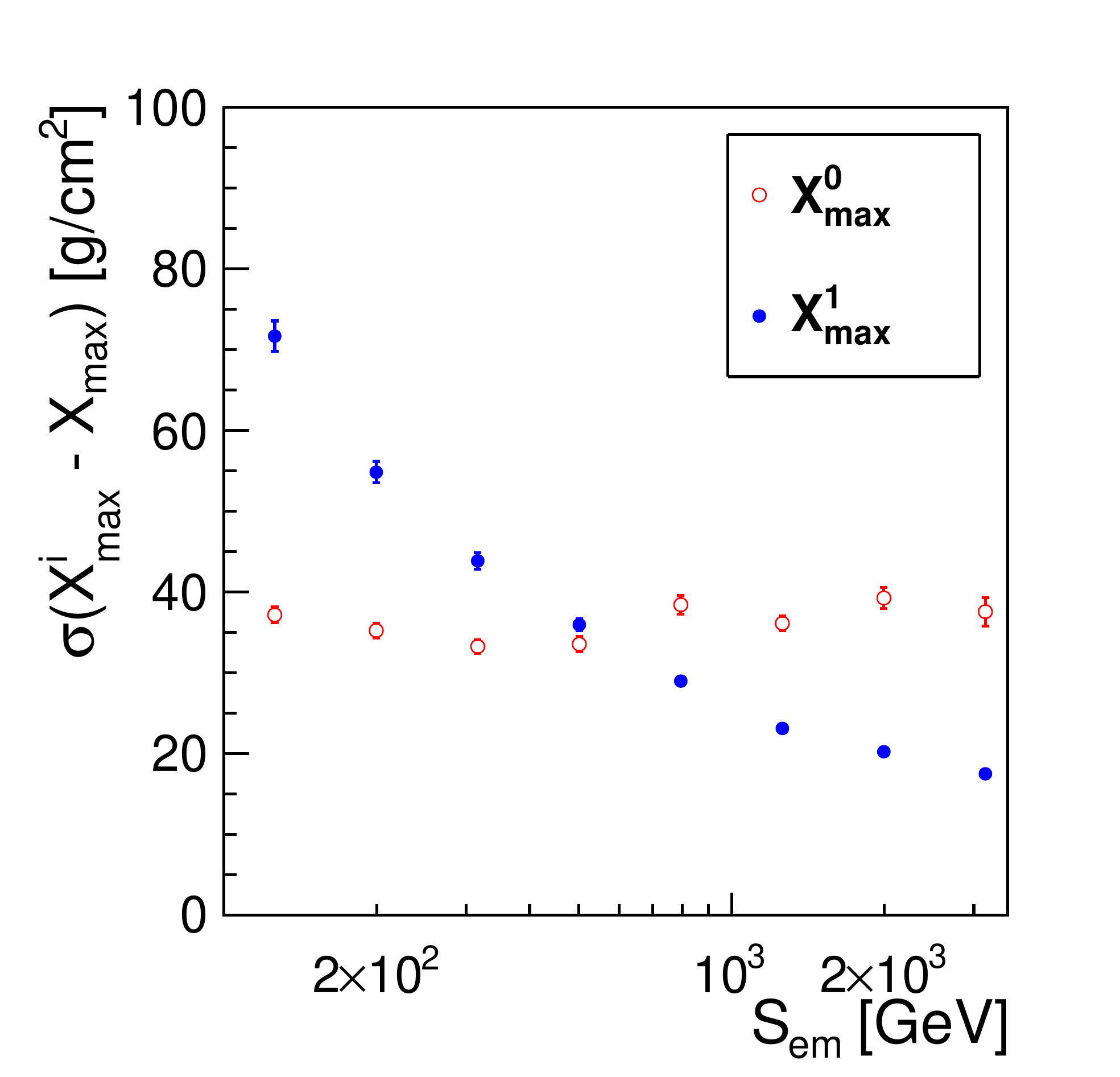}
  \caption{\label{fig:XMAX-resol} Resolution  of the \Xmax estimators \XmaxZero and \XmaxOne as a function of \Sem (see text for the simulation details).}
\end{figure}

The obtained resolutions as a function of \Sem, both for \XmaxZero and \XmaxOne, are summarized in figure \ref{fig:XMAX-resol}. Resolutions of about $40 \, {\rm g/cm^2}$ and $20 \, {\rm g/cm^2}$ were found for primaries energies of $ 1\,$TeV and  $10\,$TeV, respectively.

The two resolutions are similar in the region $A_{0}^{\rm crX} \approx 400 - 600\,$GeV.
To be on the safe side, avoiding possible tail effects, we set  $A_{0}^{\rm crX} = 600\,$GeV.
Therefore, the estimator of \Xmax, designated as $X_{\rm max}^R$ is defined as:

\begin{equation}
X_{\rm max}^R= \left\{
\begin{array}{ll}
       X_{\rm max}^1 & \ \ \ \mbox{if}\ A_{0} > A_{0}^{\rm crX}\ \\ & \ \ \ {\rm and}\  X_{\rm max}^1 > 300\,{\rm g\,cm^{-2}}\\[2pt]
      X_{\rm max}^0 & \ \ \ {\rm otherwise}\\
\end{array}
\right.
\end{equation}


\section{Energy reconstruction and resolution}
\label{sec:Erec}

In electromagnetic showers, the production of muons, either via the photo-production of mesons or by the direct creation of muon pairs is quite small \cite{muonproduction} and can thus be neglected in the global accounting of the shower energy.

\begin{figure}[!t]
  \centering
  \includegraphics[width=0.5\textwidth]{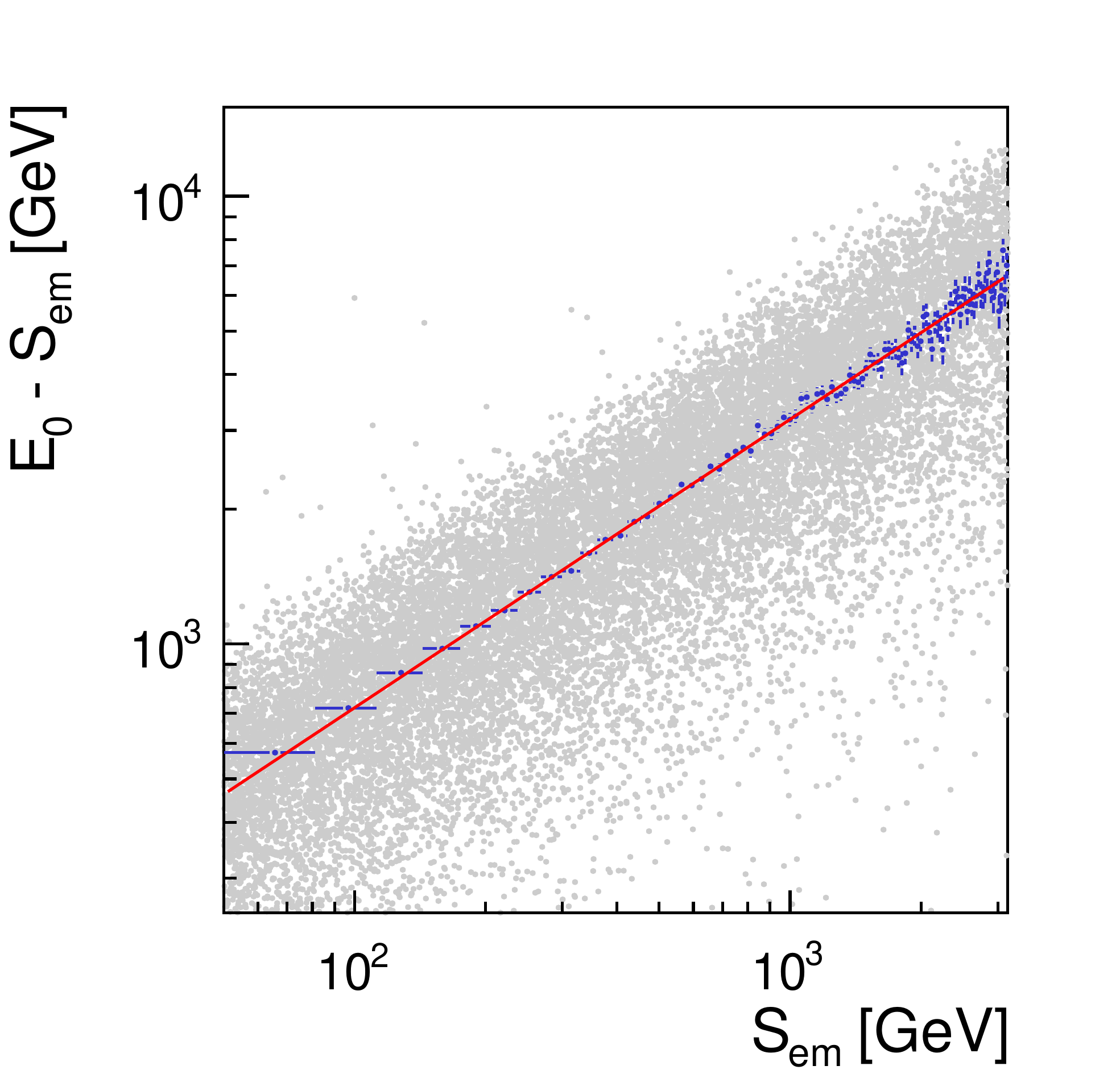}
  \caption{\label{fig:E0-SemvsSem} Correlation of the energy deposited in the atmosphere  (primary energy minus the electromagnetic energy at the ground),   $E_0-S_{\rm em}$ with the energy deposited at the ground \Sem. The line corresponds to the best parametrization (see text for details).}
\end{figure}

On the other hand, the logarithm of the electromagnetic energy deposited in the atmosphere $(E_0-S_{\rm em})$ is linearly correlated  with  the logarithm of the energy deposited at the Earth surface (\Sem), as shown in figure \ref{fig:E0-SemvsSem}.

It is then possible to parametrize $E_0$ as a function of \Sem as:

\begin{center}
\begin{equation} \label{eq:E01par}
E_{0}^{(1)} \, = \,S_{\rm em}\, +\, C \, (S_{\rm em})^{\beta} ,
\end{equation}
\end{center}

where  $C$ and $\beta$  are free positive parameters. This parametrization ensures, by construction,  that $E_{0}^{(1)}$ is always greater then \Sem. The best values found for $C$ and $\beta$  are
$ 37.2\,{\rm GeV}^{0.36}$ and $ 0.64$, respectively. The result is shown as a red curve in figure \ref{fig:E0-SemvsSem}.

Using the above parametrization and $A_0$ (see section \ref{sec:ShowerStage}) as the estimator of \Sem, it is possible to make a first energy reconstruction considering an \emph{ideal} detector\footnote{By \emph{ideal} detector it is assumed a detector able to accurately collects all the energy of electromagnetic particles reaching the station. The impact of having a \emph{real} calorimetric detector such as a water Cherenkov detector is addressed in section~\ref{sec:conclusions}.}. An energy resolution of about $40  \%$ is  obtained at $1\,$TeV.

\begin{figure}[!t]
  \centering
  \includegraphics[width=0.5\textwidth]{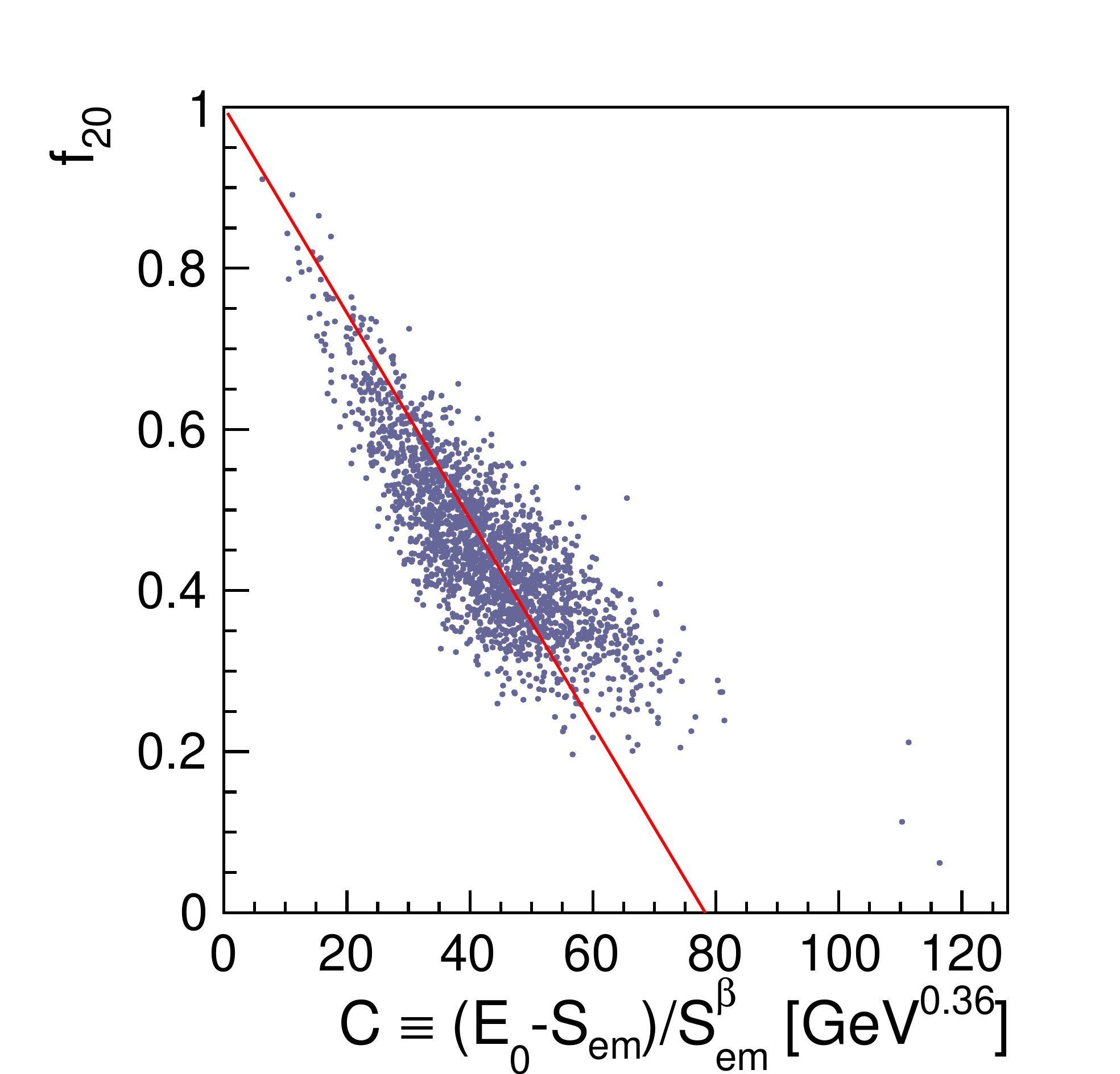}
  \caption{\label{fig:f20_C} Correlation of the variable $f_{20}$, as defined in the text,  with the calibration coefficient $C$, for events with  \mbox{\Sem$ \sim 200\,{\rm GeV}$} and $X_{\rm max} \sim 350 \,{\rm g/cm^2}$ (see text for the simulation details).}
\end{figure}

The coefficient $C$, in this first  calibration attempt, is a constant. However, $C$ can be shown to be correlated with $f_{20}$, \Xmax and \Sem.
Indeed, it is shown in figure \ref{fig:f20_C} a striking correlation between $f_{20}$ and \mbox{$ C =(E_0-S_{\rm em})/(S_{\rm em})^{\beta} $},  for events with  \Sem $\in [100;250]\,$GeV and $X_{\rm max} \in [330;385]\,\rm{g\,cm^{-2}}$.
The line in the figure is the best linear parameterization imposing that for $C= 0$ (no energy deposited in the atmosphere), $f_{20}=1$ (all the deposited at the ground is at a distance lower than 20 m from the shower core position).

\begin{figure}[!t]
  \centering
  \includegraphics[width=0.5\textwidth]{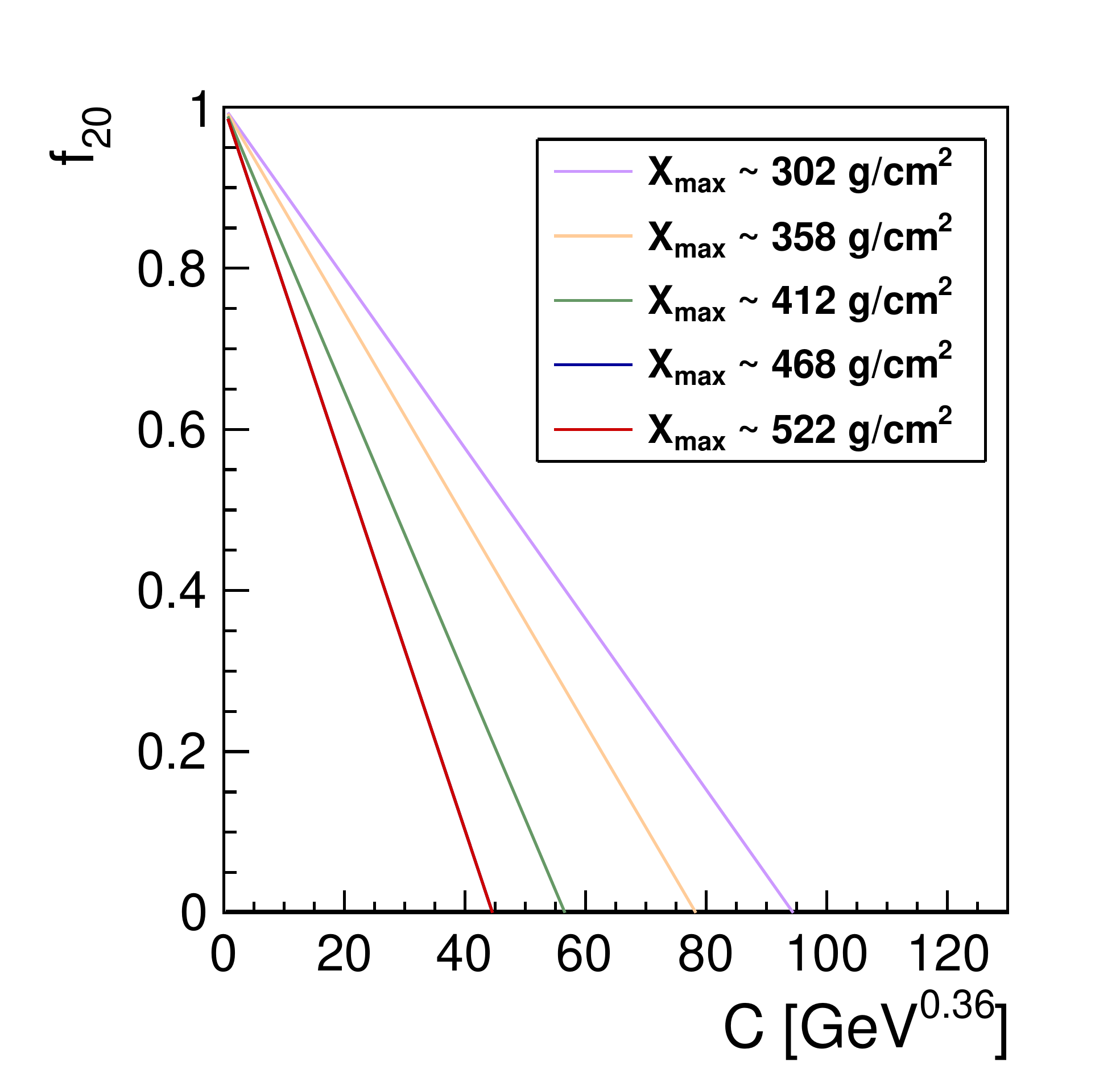}
  \caption{\label{fig:f20_E0calib} Calibration lines ($f_{20}$, $C$) for several ranges of \Xmax and with  \Sem$ \sim 200\,$GeV .}
\end{figure}

\begin{figure}[!t]
  \centering
  \includegraphics[width=0.5\textwidth]{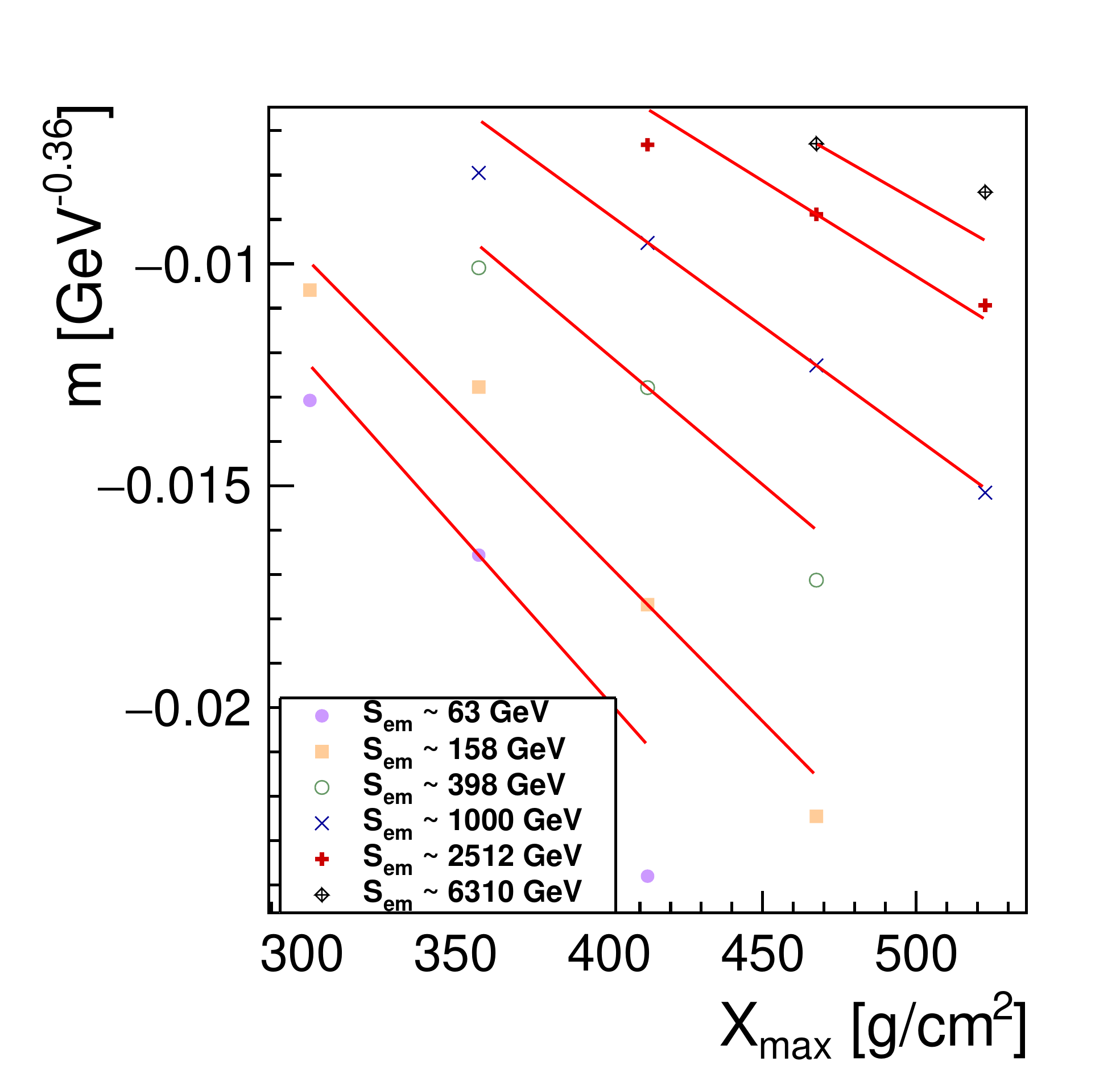}
   \caption{\label{fig:slope-f20C} Correlation  of the slope of the calibration lines represented in \ref{fig:f20_E0calib} with \Xmax, for different \Sem bins (see caption).}
\end{figure}

The set of the correlation lines ($ f_{20}, C $) for several \Xmax ranges and \Sem $\sim 200\,$GeV , are shown in figure \ref{fig:f20_E0calib}. There is a linear monotonous decrease of the slope $m$ of these lines with the increase of \Xmax. In figure \ref{fig:slope-f20C} the obtained $m$ are represented as a function of \Xmax for different bins of \Sem together with the best linear parametrization for each \Sem bin.

The extrapolation of these lines for \Xmax $=0$ points to a non-physical small positive value of $m$ \mbox{($b_m \sim 0.011\,\mathrm{GeV^{-0.36}}$)}, which means that this linear model is no longer valid for  $X_{\rm max} < 200 \, \rm{g\,cm^{-2}}$,which is far below the relevant \Xmax region for this article\footnote{For the energies considered in this article, most shower events have \Xmax values around $\sim 400 \pm 100\,{\rm g\,cm^{-2}}$. A shower with \Xmax$\gtrsim 500\,{\rm g\,cm^{-2}}$ would have its \Xmax buried in the ground while a shower with  \Xmax$\lesssim 200\,{\rm g\,cm^{-2}}$ would only reach the ground if it strongly fluctuates.}. So we will keep the linear approximation using
\mbox{$b_m = 0.011\,\mathrm{GeV^{-0.36}}$} for all \Sem.

\begin{figure}[!t]
  \centering
  \includegraphics[width=0.5\textwidth]{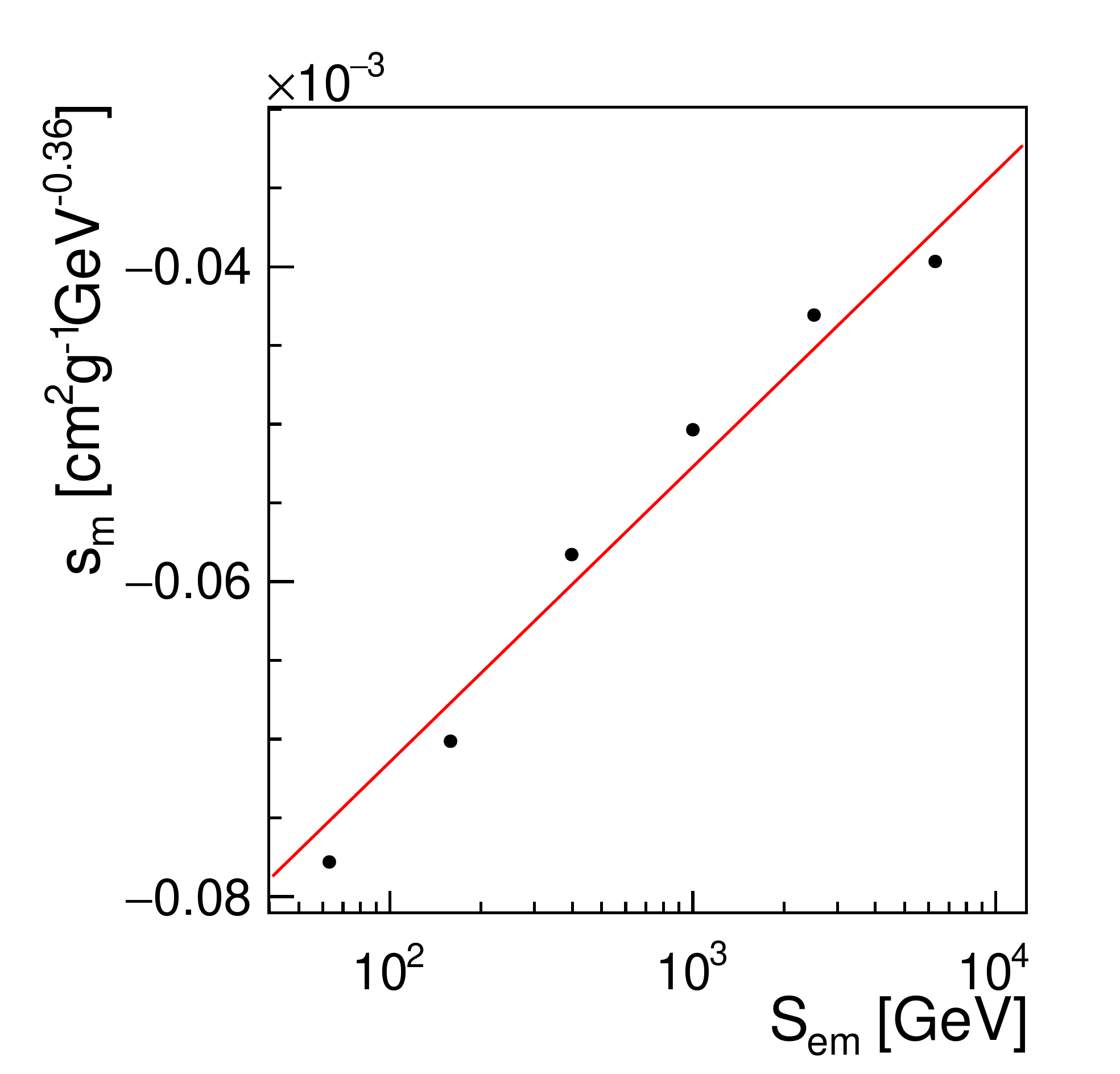}
   \caption{\label{fig:s_mSem}  Correlation of the slope $s_m$ of the $m$ calibration lines represented in fig \ref{fig:slope-f20C} as a function of \Sem.}
\end{figure}

Finally, the slope $s_m$ of the  lines represented in figure
\ref{fig:slope-f20C} are shown in figure \ref{fig:s_mSem} as a function of $\log(S_{\rm em})$. A linear correlation is found between $s_m$ and  $\log(S_{\rm em})$. As such, one can write,

\begin{center}
\begin{equation}
 f_{20} = 1 + m(X_{\rm max}, S_{\rm em}) \, C(f_{20},X_{\rm max}, S_{\rm em})\,,
\end{equation}
 \end{center}

and,

\begin{center}
\begin{equation}
 m(X_{\rm max}, S_{\rm em})= b_m + [s_{m0} + s_{m1} \,  \log(S_{\rm em}/\mathrm{GeV})] \,X_{\rm max}
\end{equation}
 \end{center}

\noindent
The best achieved parametrization with the above equation is shown in figure \ref{fig:s_mSem}, with parameters
\mbox{$s_{m0} = -1.1\times10^{-4}\, \mathrm{GeV^{-0.36} \, g^{-1} cm^{2}}$}
and
\mbox{$s_{m1} = 1.87\times10^{-5} \, \mathrm{GeV^{-0.36} \, g^{-1} cm^{2}}$}.

Finally,

\begin{center}
\begin{equation}
C(f_{20},X_{\rm max}, S_{\rm em})\,=
\frac{1-f_{20}} {- \left(b_m + [s_{m0}+s_{m1} \log(S_{\rm em})]\,X_{\rm max}\right)}
\end{equation}
\end{center}

and,

\begin{center} \label{eq:E02par}
\begin{equation}
E_{0}^{(2)} \, = \, S_{\rm em} \, +\, C (f_{20}, X_{\rm max}, S_{\rm em}) \, (S_{\rm em})^{\beta},
\end{equation}
 \end{center}

which is our best estimator for the primary energy.

\parskip 1.5ex

Using in the above parametrization, $A_0$ (see section \ref{sec:ShowerStage}) as the estimator of \Sem and $X_{\rm max}^{R}$ (see section \ref{sec:Xmax}) as the estimator of \Xmax, an energy  resolution
below $30 \, \%$ and $20 \,\%$  were obtained at $1\,$TeV and $10\,$TeV, respectively.

\parskip 0 ex

Using instead the real \Xmax and \Sem, these energy  resolutions  improve to about $8\%$ and $4\%$ . These results may be considered as the \emph{ultimate}  resolutions. The difference between the estimated resolutions and the \emph{ultimate} resolutions are driven, in the TeV region, mainly by the resolution of the \Sem estimator. Indeed, a resolution of $12\%$ on the parameter $A_0$, the \Sem estimator, even considering the simulated \Xmax value, would translate into a resolution on $E_0^{(2)}$ of $22\%$ .

\begin{figure}[!t]
  \centering
  \includegraphics[width=0.5\textwidth]{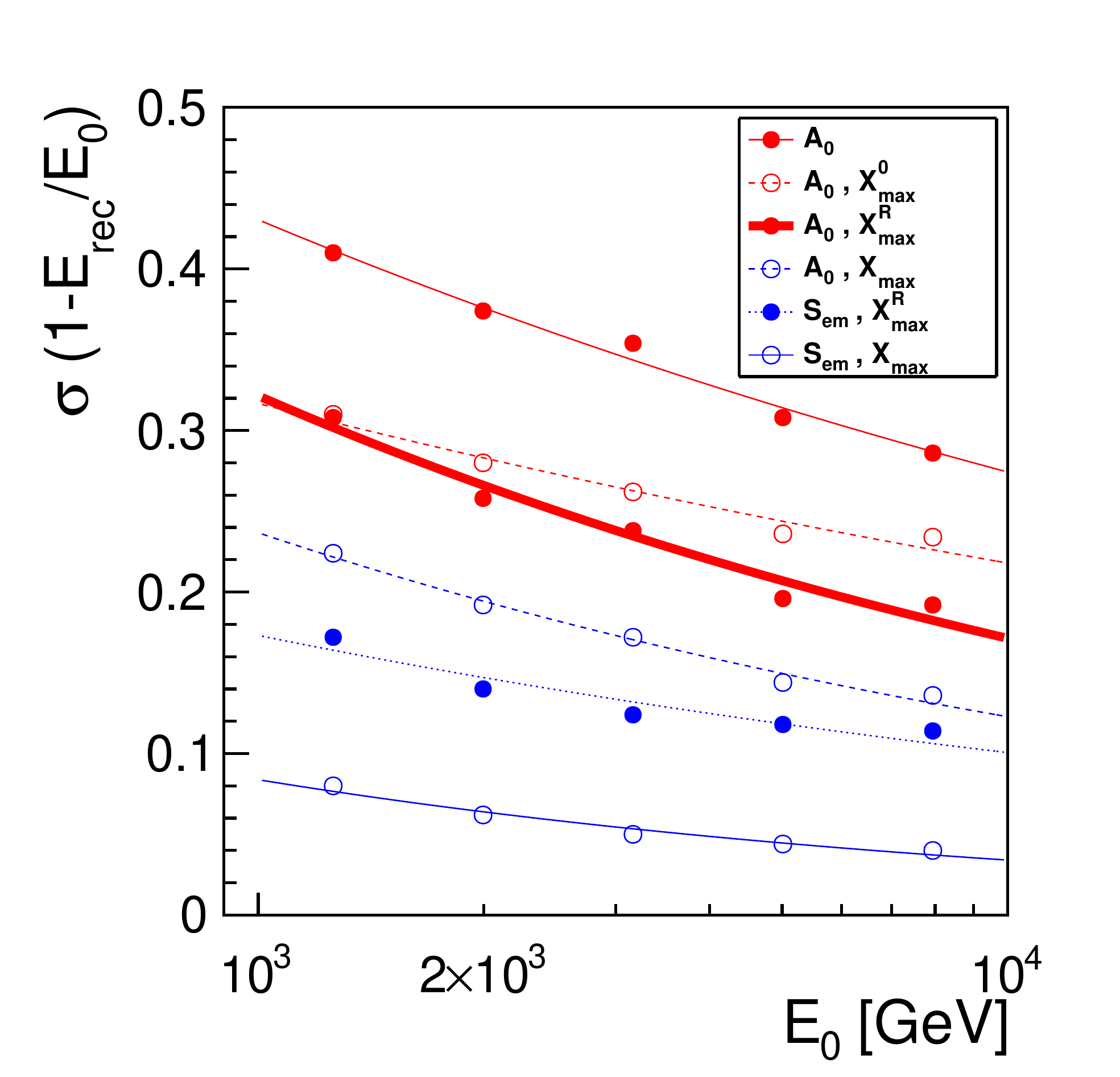}
  \caption{\label{fig:E_resol_all} Energy resolution as a function of the primary energy. The red lines correspond to constant $C$ coefficients (thin) and  $C (X_{\rm max}, f_{20}) $ coefficients, using $X_{\rm max}^{R}$  (thick) or $X_{\rm max}^{0}$ (dashed); resolutions obtained with simulated values are also shown for comparison (blue lines) : simulated value of \Xmax (dashed),  simulated value of \Sem (pointed) and both  simulated values (thin).}
\end{figure}

In figure \ref{fig:E_resol_all} are shown (full red thick line)  the estimated energy ($E_0^{(2)}$) resolution as a function of the primary energy.
For comparison, the equivalent resolutions obtained applying  the constant $C$ calibration ($E_0^{(1)}$), full red thin line, defined by equation \ref{eq:E01par}, or using systematically  the \XmaxZero estimator (dashed red thin line) are also shown.
It is also shown the resolutions that would be obtained using the simulated value of \Xmax (dashed blue line), or the simulated value of \Sem (pointed blue line) or finally using both  simulated values (blue thin line).

These results are obtained  considering an ideal detector.  The degradation factors due to the detector effects are briefly discussed in the next section.

\section{ Discussion and Conclusions}
\label{sec:conclusions}

In this article, new and innovative methods for the determination of the total electromagnetic energy at the ground, of the slant depth of the maximum of the longitudinal profile and, of the primary energy were proposed. In particular, as much as we know, it is the first time that the steepness in the core region of the cumulative function of the energy arriving at the ground is used as a determinant factor to obtain the energy calibration constants.

The obtained results are very promising and can open new physics avenues.
The improvement on the energy resolution has, as a direct consequence, a meaningful increase of the sensitivity of the observatories. The obtained resolutions on \Xmax, yet to be confirmed for cosmic ray shower, would allow very interesting studies on the development of the shower, namely on the characterization of the mass composition of hadronic cosmic rays. This shall be focus of a future publication.

These results, even if obtained considering an ideal detector, are quite robust as they rely only on the estimation of the electromagnetic energy at the ground, \Sem and, of the slant depth of the maximum of the longitudinal profile, \Xmax, which may be, in a first-order approximation, also obtained from the \Sem.  Typically, a region of a few tens of meters around the core and an energy at the ground of several tens of GeV have to be measured to be able to efficiently apply the proposed algorithms.

The estimation of \Sem is thus the critical factor to be able to achieve a good resolution on the reconstruction of the primary gamma-ray energy.
Further progress may be envisaged using more sophisticated \Sem estimators, like the one suggested by the cumulative function parametrization described in equation \ref{eq:par}, or applying, for instance, machine learning techniques.


All the present results were obtained using vertical showers. Inclined showers would imply lower energies at the ground. For instance, at $1\,$TeV an inclined $30^\circ$ shower would deposit at the ground around 20\% less than a vertical shower of the same energy (see, for instance~\cite{HAWC_GRB}). Therefore, at lower energies,  the energy resolutions will be worse, scaling with \Sem. Taking LHAASO~\cite{LHAASO_energy}, as an example, it was shown that for $10\,$TeV gamma induced showers, there is a degradation in the reconstructed energy resolution of $\sim 20\%$ when moving from showers with zenith angle $\theta \in [0^\circ,20^\circ]$ to $\theta \in [20^\circ,35^\circ]$.
It should be noted that the ability to measure higher energies may improve as the probability of the depth of shower maximum to be above the ground surface increases.

A detailed study of the performance of the new reconstruction methods for a specific Wide Field of View Gamma Ray Observatory is out of the scope of this work. Nevertheless, it is important to discuss briefly possible factors that would contribute to the degradation of the results obtained in the previous sections. Most of these effects are mainly critical for primary energies below a few hundreds of GeV, corresponding to energies at the ground of a few tens of GeV,  where a detailed design and simulation of the detectors would be mandatory.

\parskip 1.5ex

The following factors need to be considered:

\parskip 0 ex

\begin{itemize}
\item \textbf {The amount of signal (p.e.) generated in each station}

The electromagnetic energy deposited in each station may be converted into  photoelectrons (p.e.) using
as conversion factors  $n_{\rm Photon} \sim  40\, F\,{\rm p.e./GeV}$ where  $F$ is, following  \cite{WH_private}, a scale factor. $F=1$ corresponds to the existing performance at HAWC. Much higher values of $F$ were found in LATTES end-to-end simulations \cite{LATTES}. Fluctuations may then be generated using a Poisson distribution.

Anyway, for energies at $1\,$TeV or above the total signal per event is above several thousands of p.e., even for the more conservative $F$ scale factors.
The uncertainty in the measurements of $F_{50}$ and $F_{20}$ is then, at lower energies of the order of just a few \% and at higher energies negligible, which in any case will be translated at most in a minor increase of the obtained energies resolutions.

\item \textbf {The precision on the location of the shower core}

The reconstructed shower core position should, for each event,  be smeared by a few meters following the resolutions quoted in \cite{LATTES} ($3\,$m at $1\,$TeV, $ < 1 $ m at $10\,$TeV).  This would introduce distortions in the cumulative functions which would correspond to
an increase of at most a few \% in the value of the resolution of the $A$ (\Sem) estimator.

\item \textbf {The finite dimension of the compact array region of the observatory}

The distance of the core position to the border of the compact array region of the observatory
will determine the fraction of the event footprint at the ground that would be measured.
However, it is possible to imagine more sophisticated methods, using, for instance, neural networks,  to make a reasonably accurate estimation of the \Sem even in the case where there is only a partial containment of the shower footprint in the required region around the core.

\item \textbf {The time resolution at each detector station}
Time resolutions better than $2\,$ns are crucial for good angular resolutions \cite {WH_angular}. Assuming that the detectors would comply with a $2\,$ ns time resolution, the particle arrival times should be smeared and the \XmaxOne estimator (see section \ref{sec:Xmax}) re-computed.
Once again, this effect should be mainly important for lower energies; at higher energies, above a few TeV, there will be a high number of hit stations and the curvature fit will be more robust.
Nevertheless, giving up the possibility to compute, for each event,  the curvature of the shower front, the systematic use of the \XmaxZero estimator would have only impact at energies above a few TeV degrading the energy resolutions at $10\,$TeV from   $20\%$ to  $25\%$, as shown by the dashed red line in  figure~\ref{fig:E_resol_all}.
\end{itemize}

The results obtained, representing such a considerable improvement concerning the presently quoted energy resolutions of the existing or planned Wide Field of View Gamma-Ray Observatories, clearly will encourage detailed simulations and studies on the applicability of the proposed methods.

\begin{acknowledgements}
We would like to thank to A. Bueno, A. De Angelis and J. Vicha for all the useful discussions and carefully reading the manuscript.
The authors thank also for the financial support by OE - Portugal, FCT, I. P., under project PTDC/FIS-PAR/29158/2017.
R.~C.\ is grateful for the financial support by OE - Portugal, FCT, I. P., under DL57/2016/cP1330/cT0002.
\end{acknowledgements}

\bibliographystyle{spphys}       
\bibliography{References}   

\end{document}